\documentclass[preprint,showpacs,preprintnumbers,amsmath,amssymb]{revtex4}
\usepackage{graphicx}
\usepackage{dcolumn}
\usepackage{bm}

\begin{document}


\title{Continuous variable teleportation of single photon states}

\author{Toshiki Ide, Takayoshi Kobayashi}
\affiliation{%
Department of Physics, Faculty of Science, University of Tokyo,\\
7-3-1 Hongo, Bunkyo-ku, Tokyo 113-0033, Japan 
}%
 \email{ide@femto.phys.s.u-tokyo.ac.jp}
\author{Holger F. Hofmann}%
\affiliation{%
CREST, Japan Science and Technology Corporation (JST),\\
Research Institute for Electronic Science, Hokkaido University, \\
Sapporo 060-0812, Japan
}%

\author{Akira Furusawa}
\affiliation{
Department of Applied Physics, Faculty of Engeneering, University of Tokyo,\\
7-3-1 Hongo, Bunkyo-ku, Tokyo 113-8656, Japan}%

\date{August 27, 2001}

\begin{abstract}
We investigate the changes to a single photon state caused by the non-maximal entanglement in  continuous variable quantum teleportation. It is shown that the teleportation measurement introduces field coherence in the output.
\end{abstract}


\maketitle

\section{Introduction}

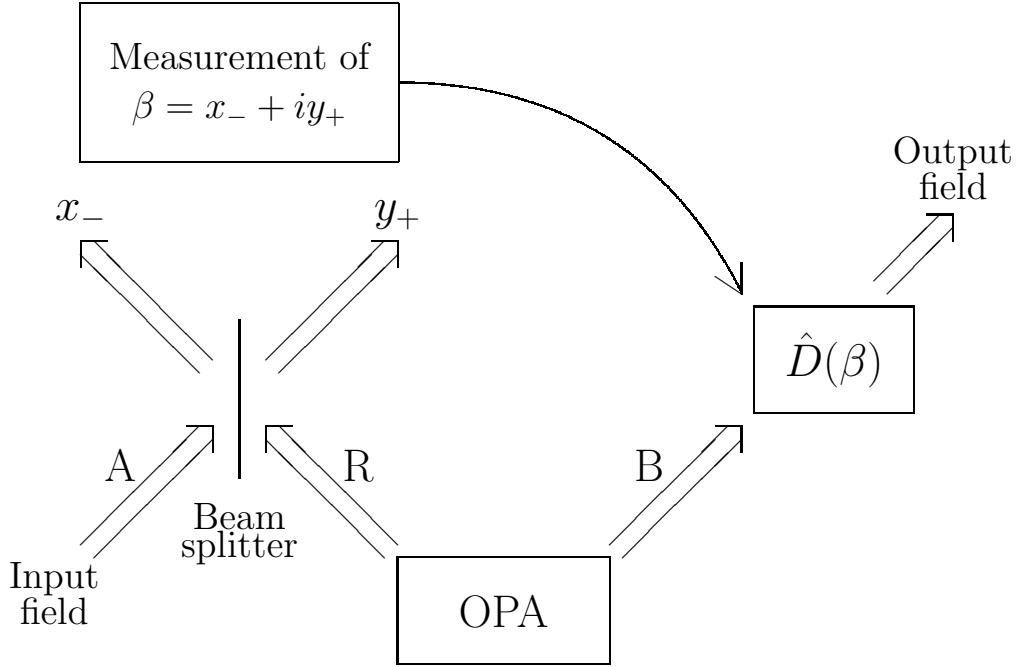
\begin{figure}
\begin{picture}(400,300)

\put(160,40){\framebox(80,40){\Large OPA}}

\put(160,85){\line(-1,1){45}}
\put(155,80){\line(-1,1){45}}
\put(110,130){\line(0,-1){10}}
\put(110,130){\line(1,0){10}}
\put(135,105){\makebox(20,20){\Large R}}

\put(240,85){\line(1,1){45}}
\put(245,80){\line(1,1){45}}
\put(290,130){\line(0,-1){10}}
\put(290,130){\line(-1,0){10}}
\put(245,105){\makebox(20,20){\Large B}}

\put(40,85){\line(1,1){45}}
\put(45,80){\line(1,1){45}}
\put(90,130){\line(0,-1){10}}
\put(90,130){\line(-1,0){10}}
\put(45,105){\makebox(20,20){\Large A}}
\put(10,66){\makebox(40,12){\large Input}}
\put(10,54){\makebox(40,12){\large field}}

\put(100,110){\line(0,1){60}}
\put(80,90){\makebox(40,12){\large Beam}} 
\put(80,78){\makebox(40,12){\large splitter}}

\put(90,155){\line(-1,1){45}}
\put(85,150){\line(-1,1){45}}
\put(40,200){\line(0,-1){10}}
\put(40,200){\line(1,0){10}}
\put(30,200){\makebox(20,20){\Large $x_-$}}

\put(110,155){\line(1,1){45}}
\put(115,150){\line(1,1){45}}
\put(160,200){\line(0,-1){10}}
\put(160,200){\line(-1,0){10}}
\put(150,200){\makebox(20,20){\Large $y_+$}}

\put(40,230){\framebox(120,60){}}
\put(60,260){\makebox(80,20){\large Measurement of}}
\put(60,240){\makebox(80,20){\large $\beta=x_-+i y_+$}}

\bezier{400}(160,260)(250,260)(290,180)
\put(290,180){\line(0,1){12}}
\put(290,180){\line(-3,2){10}}

\put(295,135){\framebox(60,40){\Large $\hat{D}(\beta)$}}

\put(340,185){\line(1,1){25}}
\put(345,180){\line(1,1){25}}
\put(370,210){\line(0,-1){10}}
\put(370,210){\line(-1,0){10}}

\put(350,227){\makebox(40,12){\large Output}}
\put(350,215){\makebox(40,12){\large field}}
\end{picture}
\caption{\label{setup} Schematic representation of the quantum teleportation 
setup.}
\label{system}
\end{figure}

Quantum teleportation is a method for Alice (sender) to transmit an unknown quantum input state to Bob (receiver) at a distant place by sending only classical information using a shared entangled state as a resource \cite{Ben93}. 
In continuous variable quantum teleportation \cite{Vai94,Brau98,Fur98}, the available entanglement is non-maximal, limited by the amount of squeezing achieved. Fig.\ref{setup} shows the setup of a continuous variable quantum teleportation. Alice transmits an unknown quantum state $\mid \psi \rangle _{A}$ to Bob. Alice and Bob share EPR beams in advance. 
Alice mixes her input state with the reference EPR beam by a 50$\%$ beam-splitter and performs an entanglement measurement of the complex field value $\beta$.
After Bob gets the information of the field measurement value $\beta$ from Alice, Bob applies a displacement to the output state by mixing the coherent field of a local oscillator with the output EPR beam $B$. 

As has been shown previously \cite{Hof00}, the properties of this transfer process can be summarized by the transfer operator $\hat{T}_q(\beta)$ which describes both the probability distribution $P(\beta)$ of measurement results $\beta$ and the normalized conditional output state \\
$\mid \psi_{\mbox{out}}(\beta)\rangle$ 
for any input state $\mid \psi_{\mbox{in}}\rangle$, such that
\begin{equation}
\sqrt{P(\beta)} \mid \psi_{\mbox{out}}(\beta)\rangle = \hat{T}_q(\beta) \mid \psi_{\mbox{in}}\rangle.
\end{equation}
In its diagonalized form, this transfer operator reads 
\begin{equation}
\hat{T}_{q}(\beta) = \sqrt{\frac{1-q^2}{\pi}}\sum_{n=0}^{\infty}q^n \hat{D}(\beta) \mid n \rangle _{BA}\langle n \mid \hat{D}(-\beta).
\label{transfer}
\end{equation}
The non-maximal entanglement is described by the parameter $q$, which is 0 for a non-entangled vacuum and 1 for maximal entanglement. In the following, this operator will be applied to characterize the teleportation of a single photon input state, with special consideration of the field coherence created in the output by the teleportation process. 

\section{Teleportation of a single photon state}
The output of a one photon input state is characterized by
\begin{eqnarray}
\hat{T}_q(\beta) \mid 1 \rangle = \sqrt{\frac{1-q^2}{\pi}}
e^{-(1-q^2)\frac{|\beta|^2}{2}} 
\hat{D}((1-q)\beta) 
((1-q^2)\beta^* \mid 0 \rangle + q \mid 1 \rangle).
\label{oneout}
\end{eqnarray}
The normalized output state is then given by
\begin{eqnarray}
\mid \psi_{\mbox{out}}(\beta)\rangle 
&=& 
\frac{1}{\sqrt{q^2+(1-q^2)^2 |\beta|^2}} 
\hat{D}((1-q)\beta)((1-q^2)\beta^* \mid 0 \rangle + q \mid 1 \rangle).
\end{eqnarray}
The output state can be described by a displaced coherent 
superposition of a zero photon and a one photon component. 
Both the displacement and the coherence depend on the 
complex measurement value $\beta$. They therefore represent
a measurement induced coherence of the output state. We 
characterize this coherence by the expectation value of the
 complex field amplitude,
\begin{eqnarray}
C_q(\beta) &=& \langle \psi_{\mbox{out}}(\beta) \mid \hat{a} 
\mid \psi_{\mbox{out}}(\beta)\rangle \nonumber \\
&=& \frac{q (1-q^2)}{q^2+(1-q^2)^2 |\beta|^2} \beta +
(1-q)\beta.
\end{eqnarray}

\begin{figure}[htbp]
\begin{picture}(350,400)
\put(50,200){\makebox(70,20){\Large $C^{(1)}_{\frac{1}{2}}(|\beta|)$}}
\put(40,0){\makebox(300,200){\includegraphics[width=10cm]{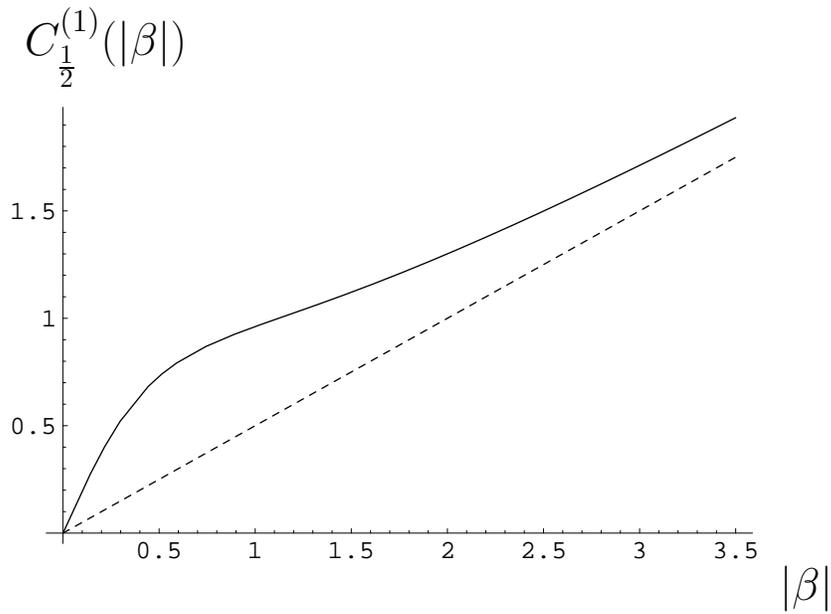}}}
\put(330,0){\makebox(40,10){\Large $|\beta|$}}
\end{picture}
\caption{The absolute value of $\beta$ dependence of the field amplitude $C^{(1)}_{q}(\beta)$ in the case of $q=0.5$. The curve is approaching $y=0.5q$.}
\label{coh_b}
\end{figure}
Fig.\ref{coh_b} illustrates this dependence of $C_q$ on $\beta$ for
  $q=1/2$. For low values of $\beta$, $C_q$ rises sharply,
  levelling off around $|\beta|\approx 0.5$. At higher values
  of $|\beta|$, $C_q$ slowly approaches $(1-q)\beta$, as 
  indicated by the dotted line. 
  Since the input photon number state has a field expectation
  value of zero, the field coherence in the output is a 
  consequence of the measurement $\beta$. As has been argued
  elsewere \cite{Hof00}, $\beta$ corresponds to the result of a 
  field measurement performed on the input state. This
measurement creates coherence by projection onto displaced photon number
 states as indicated by the diagonalized form of $\hat{T}_q(\beta)$ in equation (\ref{transfer}). Since the photon
  is the quantum mechanical equivalent of field intensity, each
  photon represents an addition of one quantum unit to the field 
  fluctuations. The measurement then converts the field
fluctuation
  into an actual field. This process is responsible for the
  rapid increase in coherence at low values of $\beta$. 

\section{Conclusions}
We have investigated the effects of continous variable quantum teleportation on a single photon input state. Because of the non-maximal entanglement used in the teleportation, the measurement of $\beta$ introduces coherence into the output state. We have quantified this coherence, tracing its origin to the field fluctuations of the single photon input.

\end{document}